\documentclass[11pt]{article}
\usepackage{amsmath}
\usepackage{amssymb}
\usepackage{amsthm,amsxtra}
\usepackage{epsf}
\usepackage{eepic}
\newlength{\mytopmargin}
\newlength{\myleftmargin}
\newtheorem{prop}{Proposition}
\newcommand{\zz}{\mathbb Z}

\newcommand\psymmU{%
\begin{picture}(1,1)(0,0)%
\allinethickness{0.5pt}%
\path(0,0)(0,1)(1,1)(1,0)(0,0)%
\end{picture}}
\newcommand\psymmUU{%
\begin{picture}(1,1)(0,0)%
\allinethickness{0.5pt}%
\path(0,0)(0,1)(1,1)(1,0)(0,0)%
\put(0.5,0.5){\makebox(0,0){$\cdot$}}%
\end{picture}}
\newcommand\psymmO{%
\begin{picture}(1,1)(0,0)%
\allinethickness{0.5pt}%
\path(0,0)(0,1)(1,1)(1,0)(0,0)%
\path(0,0)(1,1)%
\end{picture}}
\newcommand\psymmS{%
\begin{picture}(1,1)(0,0)%
\allinethickness{0.5pt}%
\path(0,0)(0,1)(1,1)(1,0)(0,0)%
\path(1,0)(0,1)%
\end{picture}}
\newcommand\psymmu{%
\begin{picture}(1,1)(0,0)%
\allinethickness{0.5pt}%
\path(0,0)(0,1)(1,1)(1,0)(0,0)%
\path(0,0)(1,1)%
\path(0,1)(1,0)%
\end{picture}}

\newbox\tsymmUbox
\newbox\tsymmUUbox
\newbox\tsymmObox
\newbox\tsymmSbox
\newbox\tsymmubox
\setbox\tsymmUbox =\hbox{\kern0.75pt\setlength{\unitlength}{6pt}\psymmU \kern0.75pt}

\setbox\tsymmUUbox=\hbox{\kern0.75pt\setlength{\unitlength}{6pt}\psymmUU\kern0.75pt}
\setbox\tsymmObox =\hbox{\kern0.75pt\setlength{\unitlength}{6pt}\psymmO \kern0.75pt}
\setbox\tsymmSbox =\hbox{\kern0.75pt\setlength{\unitlength}{6pt}\psymmS \kern0.75pt}
\setbox\tsymmubox =\hbox{\kern0.75pt\setlength{\unitlength}{6pt}\psymmu \kern0.75pt}

\newbox\symmUbox
\newbox\symmUUbox
\newbox\symmObox
\newbox\symmSbox
\newbox\symmubox
\setbox\symmUbox =\hbox{\kern0.75pt\setlength{\unitlength}{4.5pt}\psymmU \kern0.75pt}
\setbox\symmUUbox=\hbox{\kern0.75pt\setlength{\unitlength}{4.5pt}\psymmUU\kern0.75pt}
\setbox\symmObox =\hbox{\kern0.75pt\setlength{\unitlength}{4.5pt}\psymmO \kern0.75pt}
\setbox\symmSbox =\hbox{\kern0.75pt\setlength{\unitlength}{4.5pt}\psymmS \kern0.75pt}
\setbox\symmubox =\hbox{\kern0.75pt\setlength{\unitlength}{4.5pt}\psymmu \kern0.75pt}
\def\symmU{{\copy\symmUbox}}
\def\symmUU{{\copy\symmUUbox}}
\def\symmO{{\copy\symmObox}}
\def\symmS{{\copy\symmSbox}}
\def\symmu{{\copy\symmubox}}

\begin{document}

\vspace{4cm}
\noindent
{\bf Growth models, random matrices and Painlev\'e transcendents}

\vspace{5mm}
\noindent
Peter J.~Forrester${}^*$\footnote{Supported by the Australian Research
Council}

\noindent
${}^*$Department of Mathematics and Statistics,
University of Melbourne, \\
Victoria 3010, Australia 
\begin{quote}
The Hammersley process relates to the statistical properties of the
maximum length of all up/right paths connecting random points of a
given density in the unit square from (0,0) to (1,1). This process can
also be interpreted in terms of the height of the polynuclear growth
model, or the length of the longest increasing subsequence in a random
permutation. The cumulative distribution of the longest path length can
be written in terms of an average over the unitary group. Versions of the
Hammersley process in which the points are constrained to have certain
symmetries of the square allow similar formulas. The derivation of these
formulas is reviewed. Generalizing the original model to have point
sources along two boundaries of the square, and appropriately scaling
the parameters gives a model in the KPZ universality class. Following
works of Baik and Rains, and Pr\"ahofer and Spohn, we review the
calculation of the scaled cumulative distribution, in which a particular
Painlev\'e II transcendent plays a prominent role.
\end{quote}

\section{Introduction}
The aim of this review is to explain aspects
of developments over the past few years
relating some observables in statistical mechanics models to random matrix
averages and then to Painlev\'e transcendents. In addition to the
theoretical interest in these inter-relationships, the fact that the
Painlev\'e transcendents are readily computable means that quantitative
predictions for certain order one scaled observables are available for the
first time. Perhaps the most significant such result is the calculation due
to Pr\"ahofer and Spohn \cite{PS02} of the exact two-point scaling function for
one-dimensional stationary KPZ (Kardar-Parisi-Zhang) growth. KPZ growth
is generally believed (see e.g.~\cite{HHZ95,BS95}) 
to underlie a diverse number of
growth models in $1+1$ dimension. One model within the KPZ universality is
the polynuclear growth (PNG) model. The particular variant of the latter
relevant in this context was solved in terms of a Painlev\'e transcendent
known from random matrix theory \cite{TW94a} by Baik and Rains \cite{BR01a},
and it is this solution which is interpreted and computed in
\cite{PS02}. Section 4 of the present work gives some details of the exact
solution.

We begin in Section 2 by reviewing the calculation of the cumulative 
distribution for the longest path in the Hammersley process. This will be
shown to be equivalent to computing the cumulative distribution for the
maximum height in the PNG model, the longest increasing subsequence length
for a random permutation, or the maximum displacement of certain families of
non-intersecting paths. The cumulative distribution is given as a particular
random matrix average over the unitary group. In Section 3, four different
symmetrizations of the Hammersley process are considered. The
cumulative distributions of the longest path in each of these cases can
again be written as particular random matrix averages, involving the
orthogonal and symplectic groups in two of the cases, and the unitary group
in the remaining two. After presenting some details of the calculation
of the scaled distribution for the variant of the PNG model of relevance
to KPZ growth in Section 4, we conclude in Section 5 by indicating aspects of 
the Painlev\'e transcendent content of  the averages over the orthogonal and
symplectic groups encountered in Section 2.

\section{The Hammersley process}
\setcounter{equation}{0}
\subsection{Relationship to permutations}
\begin{figure}[th]
\epsfxsize=4.5cm
\centerline{\epsfbox{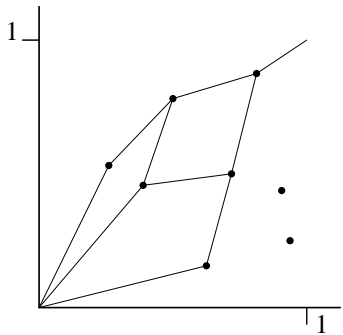}}
\caption{\label{NL.f1} Eight points in the unit square, and the extended
directed paths of maximum length. Since the number of points in these
paths equals three, here $l^\symmU_n = 3$.
}
\end{figure}

The Hammersley process (see \cite{AD95,AD99} for an extended account 
of different emphasis to that given here,
and
for references to the original literature) refers to the
following stochastic model. In the unit square mark in points uniformly
at random according to a Poisson rate with intensity $\lambda^2$, so that
the probability the square contains $N$ points  is equal to
$\lambda^{2N} e^{-\lambda^2 } / N!$. Form a continuous path by joining points
with straight line segments of positive slope, which are thus
orientated up and to the right. Extend this path to
begin at $(0,0)$ and finish at $(1,1)$ by adding an extra segment at
both ends, and define the length of the extended path as the number of points
it contains. 
Take as the primary observable quantity the stochastic variable,
$l^\symmU$ say, specifying the maximum of the lengths of all possible
extended paths (see Figure \ref{NL.f1}).

For any particular realization of exactly $N$ points the  Hammersley process
gives a geometrical construction of a random permutation of
$\{1,2,\dots,N\}$. This comes about by first labelling the $x$
coordinates of the points by $0<x_1<\cdots<x_N<1$ and similarly the
$y$ coordinates by $0<y_1<\cdots<y_N<1$. 
Each point will then have a
coordinate of the form $(x_j, y_{P(j)})$ where $\{P(1),\dots,P(N)\}$
is a permutation of $\{1,2,\dots,N\}$. The quantity $l^\symmU$ also has
an interpretation in terms of the permutation. Thus the analogue of an 
up/right path
connecting points is a subsequence
$1 \le j_1 < j_2 < \cdots < j_r \le N$ such that
$P(j_1) < P(j_2) < \cdots < P(j_r)$, which is referred to as an
increasing subsequence. The length of an increasing subsequence is defined as
the value $r$. We then see from the definitions that the maximum
length of all increasing subsequences of $P$ coincides with $l^\symmU$.

\subsection{Polynuclear growth model}
\begin{figure}[th]
\epsfxsize=7cm
\centerline{\epsfbox{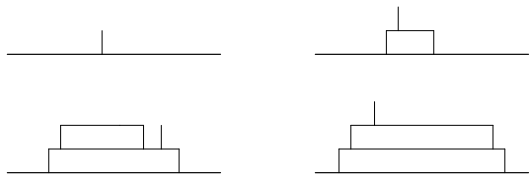}}
\caption{\label{fNL.3} Example of the plateau profile at the time of four
successive nucleation events, including the initial event (which is
labelled the 0th event and its plateau the 0th level). Note that between
the second and third nucleation event, two plateaus on the first
level have coalesced.
}
\end{figure}

Consider the $x$-$t$ half plane $t>0$. Let this half plane be filled with
points uniformly at random and such that the mean density is unity.
These points are to be thought of as seeds for nucleation events of
layered growth. In a droplet model, at $(x,t)=(0,0)$ a single layer, taken
to have zero height, starts spreading with unit velocity to the left and
to the right. Forming on top of the ground layer are new layers of unit
height. These layers, or parts thereof, are formed at space-time positions
$(x_i,t_i)$ for each nucleation event bounded by
the `lightcone' axis $u=(t+x)/\sqrt{2}$, $v=(t-x)/\sqrt{2}$;
nucleation events outside this cone are not created at a time that their
position coordinate makes contact with the ground layer or its growth. The
nucleation events $(x_i,t_i)$ inside the lightcone create the beginning of
a portion of a layer of unit height on top of the ground layer, or existing
layers, at position $x_i$. The layers are formed by the growth of the
nucleation events with unit velocity to the left and to the right; if two
growing portions of a layer collide, then growth at that point ceases and
the two portions become one, growing only at the end points of this one
portion (see Figure \ref{fNL.3} for an example). Of interest is the
statistical properties of the height at the origin after this growth
process --- known as the polynuclear growth (PNG) model --- has been
underway for time $t=T$.

\vspace{.5cm}

\begin{figure}[th]
\epsfxsize=8cm
\centerline{\epsfbox{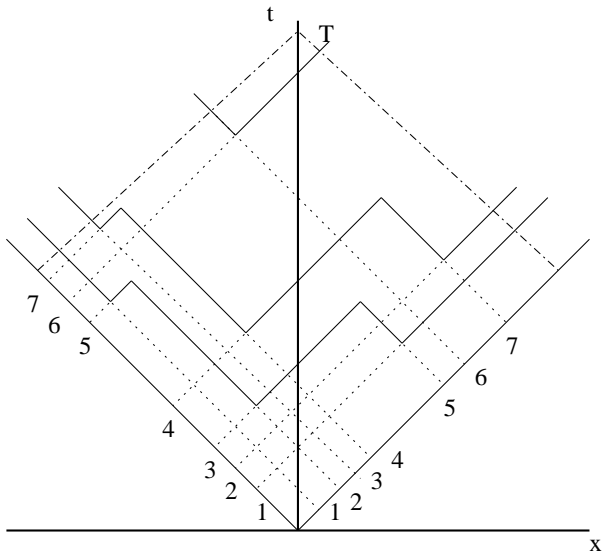}}
\caption{\label{fNL.4} World lines for the endpoints of the plateaux.
Only nucleation events inside the square shaped region including the lines
$x = \pm t$, and the lines from $t = T$ to these lines affect the height
at the origin.
The nucleation points occur at v shaped configurations, while
the inverted v part of the worldlines correspond to
the joining of the plateaux originating from two different nucleation
events. The labelling on the lines $x=t$ and $x=-t$ allow the world
lines to be identified uniquely with a permutation.
}
\end{figure}

The first observation is that only those nucleation events in the region
$[u=0,u=T/\sqrt{2}] \times 
[v=0,v=T/\sqrt{2}]$ of the lightcone can contribute to the height at
$x=0$ up to time $t=T$. Suppose in a realization of the nucleation events
there are $N$ points in this region. For a Poisson process of unit density
this occurs with probability 
$\lambda^{2N} e^{-\lambda^2}/N!$, where $\lambda^2 = T^2/2$ is the area of the
region. Use the construction of the previous subsection to associate with
the configuration of points a permutation $P$
(see Figure \ref{fNL.4}). Also indicated in Figure \ref{fNL.4} are the
world lines of the nucleation events, which show clearly the layered
structure of the growth, and in particular the height at the origin
after time $T$. The layers in which the particular nucleation events
occur are simply related to the permutation $P$. This is done by
partitioning the permutation into decreasing subsequences using the
leftmost digits at all times. The $j$th such decreasing subsequence
corresponds to the $j$th layer in the growth process. For example, in
Figure \ref{fNL.4} the permutation is 5374162, and the decreasing
subsequences formed from the leftmost digits are $(531)(742)(6)$. It
is easy to see that in general the number of decreasing subsequences of
this type is equal to the length of the longest increasing subsequence
of the same permutation. Thus studying the height at the origin in the
PNG model after time $T$ is equivalent to studying the maximum path
length in the Hammersley process with intensity $\lambda^2 = T^2/2$.

\vspace{.5cm}

\subsection{Robinson-Schensted-Knuth correspondence}
To analyze the Hammersley process requires a formula for the number of
up/right paths 
less than some prescribed value $l$ say. For this
purpose, in a realization containing $N$ points, we first associate with
each point $(x_j,y_{P(j)})$, $j=1,\dots,N$ the permutation matrix defined
so that the entries $(j,P(j))$ are equal to unity with all other entries
equal to zero. We then apply what is essentially Viennot's shadow method
\cite{Vi77} to give a bijection between permutation matrices and
certain configurations of lattice paths, the outermost member of which
can be interpreted as the profile of a lattice variant of the PNG model
\cite{Jo02}.

In fact it is possible to give a bijection between a $n \times n$
non-negative integer matrix $X=[x_{i,j}]_{i,j=1,\dots,n}$ and a pair
of so called up/ right horizontal (u/rh) non-intersecting lattice
paths. This bijection is equivalent to the celebrated 
Robinson-Schensted-Knuth correspondence mapping $X$ to a pair of semi-standard
tableaux \cite{Fu97}. The non-intersecting lattice paths
are defined on the square lattice and start at
$x=0$, one unit apart in the $y$-direction at $y=0,\dots,-(n-1)$,
and finish at $x=n-1$, with $y$-coordinates $\mu_l - (l-1)$
$(l=1,\dots,n)$ where $\mu_1 \ge \mu_2 \ge \cdots \ge \mu_N \ge 0$.
The path starting at $y=-(l-1)$ is referred to as the level-$l$ path. 
Each path may move either up or to the right along the edges of the
lattice, with the constraint that the paths may not intersect. When
$X$ is a permutation matrix and so has exactly one non-zero entry,
equal to unity, in each row and column the paths are further restricted
so that for each allowed $x$-value there is exactly one up step in total,
which is of unit length.

\begin{figure}[th]
\epsfxsize=8cm
\centerline{\epsfbox{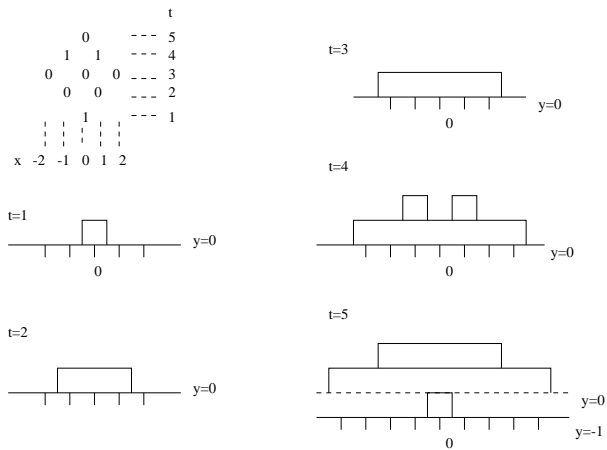}}
\caption{\label{NLR.f1} The RSK mapping in the non-intersecting
paths picture from a permutation matrix to a pair of u/rh 
non-intersecting paths, constrained so that there is
exactly one up set in total for each allowed value of $x$.} 
\end{figure}

In the general case the entries $x_{i,j}$ of $X$, where for convenience
the rows are labelled from the bottom, represent the heights of columns
of unit length centred about $x=j-i$ which occur at time $t=i+j-1$.
This labelling is simple to implement
by first rotating the matrix $45^\circ$ anti-clockwise. The columns are to
be placed on top of the level-1 path formed by earlier nucleation events
and their growth. On this latter point, during each time interval the
existing profile or profiles are required to grow one unit to the left
or to the right, with any resulting overlap recorded on the path one
level below. What results are a set of at most $n$ non-intersecting lattice
paths, equivalent to a pair of u/rh non-intersecting lattice paths each
with the same final positions. Furthermore, it is easy to see that the
process is invertible, and so there is a bijection between 
non-negative integer matrices and pairs of u/rh non-intersecting
lattice paths. As already remarked, in the case of permutation matrices,
these paths have the additional constraint of containing exactly one up
step in total for each allowed $x$-value, 
$x = \pm (2n+3/2-2j)$ ($j=1,\dots,n$).
An example is given in
Figure \ref{NLR.f1}.

Non-intersecting u/rh lattice paths, constrained to have exactly one
up step in total for each allowed $x$-value can be encoded as standard
tableaux. The latter consist of an array of unit boxes stacked across
rows of length $\mu_1, \mu_2,\dots,\mu_n$ respectively with the first
box in each row contained in the first column of the array etc. The
standard tableau is said to have shape $\mu$. One plus the value of
the $x$-coordinates that contain up steps are marked in order along
row $l$. Thus the numbers in the array have the property of being
strictly increasing both along rows and down columns, with each number
from $\{1,2,\dots,n\}$ recorded once. The standard tableau is said to have
content $n$.

\begin{figure}[th]
\epsfxsize=7cm
\centerline{\epsfbox{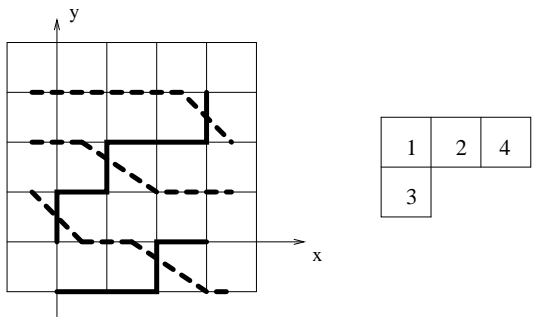}}
\caption{\label{fNL6.fig} 
Drawn in heavy lines on the square lattice is a family
of two u/rh lattice paths, constrained so that there is exactly one up step
in total for each $x$-value, and with maximum displacement 3 units, while
drawn in heavy dashed lines are the corresponding three dual lattice paths. 
Also given is the standard tableau encoding the two 
original u/rh lattice paths.}
\end{figure}

Let $f_N^\mu$ denote the number of standard tableaux of shape $\mu$ and
content $N$. Let $l_N^\symmU$ denote the longest increasing subsequence
length in a realization of the Hammersley process containing $N$ points. Then
it follows from the correspondence with pairs of constrained u/rh lattice
paths and thus pairs of standard tableau that
\begin{equation}\label{4.1}
{\rm Pr}(l_N^\symmU \le l) = {1 \over N!} \sum_{\mu: \mu_1 \le l} (f_N^\mu)^2.
\end{equation}
As a first step towards evaluating the sum in (\ref{4.1}), note that the
symmetry between rows and columns in a standard tableau implies we
can write
\begin{equation}\label{5.1}
{\rm Pr}(l_N^\symmU \le l) = {1 \over N!} \sum_{\mu: \mu_1' \le l} (f_N^\mu)^2
\end{equation}
where $\mu_1'$ denotes the length of the first column. In terms of lattice
paths, this can be understood from the bijection between u/rh lattice
paths and dual u/rh lattice paths, where if in the former the maximum
displacement is $l$, then in the latter there are exactly $l$ lattice
paths (see Figure \ref{fNL6.fig}).

\subsection{Relationship to the lock step model of vicious walkers}
The task of computing the number of pairs of constrained u/rh lattice
paths, each with the same final positions, and containing exactly
$l$ lattice paths can be considered as a special case of a more general
counting problem. First we remark that a set of $l^* < l$ lattice
paths between $x=0$ to $x=n-1$ can be uniquely converted to a set of
$l$ lattice paths by drawing horizontal segments from 
$x=0$ to $x=n-1$ along $y=-(l^*-1),\dots,-(l-1)$, so to compute
(\ref{5.1}) the case of precisely $l$ lattice paths is what is relevant.
We generalize the rules of constructing the lattice paths so that for
each allowed $x$ value $x=0,1,\dots,n-1$ there is again only one segment
which is not horizontal, but now this segment may be either up or down,
subject again to the non-intersecting constraint. These paths can be
considered as the world lines for the stochastic evolution of $l$
random walkers on a one-dimensional lattice. At each tick of the clock
exactly one walker may move to the left or to the right one lattice
space, subject to the constraint that this lattice site is not already
occupied. This defines what is known as the random turns model of
vicious walkers \cite{Fi84,Fo90b}. Let an up (down) step at $x=j-1$ be
weighted $w_j^+$ ($w_j^-$), and suppose furthermore that each level-$k$
path returns to $y=-(k-1)$ at $x=n-1$. Then with
$$
G_{l,n} := \sum_{\rm paths} \prod_{k=1}^l( {\rm weight \: of \:
level}\mbox{-}k \: {\rm path})
$$
it is shown in \cite{Fo90b,Fo01a} that
\begin{equation}\label{6.1}
G_{l,n} = {1 \over (2 \pi)^l l!}
\int_{-\pi}^\pi d \theta_1 \cdots  \int_{-\pi}^\pi d \theta_l \,
\prod_{k=1}^n \sum_{j=1}^l (w_k^+ e^{i \theta_j} +
w_k^- e^{-i \theta_j})
\prod_{1 \le j < k \le l} | e^{i \theta_k} - e^{i \theta_j} |^2.
\end{equation}
But 
$$
\sum_{\mu: \mu_1' \le l} (f_N^\mu)^2 = G_{l,2N}
\Big |_{w_p^-=0, \: w_p^+=1 \: (p=1,\dots,N) \atop
w_p^+=0, \: w_p^-=1 \: (p=N+1,\dots,2N)}
$$
so we see from (\ref{5.1}) and (\ref{6.1}) that
\begin{eqnarray}\label{27.1}
{\rm Pr}(l_N^\symmU \le l) & = & {1 \over N!} {1 \over (2 \pi)^l l!}
\int_{-\pi}^\pi d \theta_1 \cdots  \int_{-\pi}^\pi d \theta_l \,
\Big | \sum_{j=1}^l \exp{i \theta_j}  \Big |^{2N}
\prod_{1 \le j < k \le l} | e^{i \theta_k} - e^{i \theta_j} |^2
\nonumber \\
& = & {N! \over (2N)!}
\Big \langle \Big (\sum_{j=1}^l 2 \cos \theta_j \Big )^{2N}
\Big \rangle_{U(l)}
\end{eqnarray}
where $\langle \cdot \rangle_{U(l)}$ 
denotes an average over the eigenvalue 
probability density function for 
random matrices from the classical group $U(l)$,
$$
{1 \over (2 \pi)^l l!} \prod_{1 \le j < k \le l} 
| e^{i \theta_k} - e^{i \theta_j} |^2, \qquad - \pi < \theta_j < \pi.
$$
To see the validity of the second equality in (\ref{27.1}), one first notes
$$
 \Big (\sum_{j=1}^l 2 \cos \theta_j \Big )^{2N} =
\sum_{p=0}^{2N} \Big ( {2N \atop p} \Big )
\Big ( \sum_{j=1}^l e^{i \theta_j} \Big )^p
\Big ( \sum_{j=1}^l e^{-i \theta_j} \Big )^{2N-p} 
$$
and then observes that only the $p=N$ term in this sum is non-zero after
averaging over $U(l)$.
Now, according to the
definitions
$$
{\rm Pr}(l^\symmU \le l) = e^{-\lambda^2 } \sum_{N=0} {\lambda^{2N} \over
N!} {\rm Pr}(l_N^\symmU \le l)
$$
so substituting (\ref{27.1}) gives \cite{Ge90,Ra98}
\begin{equation}\label{27.2}
{\rm Pr}(l^\symmU \le l) = e^{-\lambda^2 }
\Big \langle e^{{\lambda} {\rm Tr} (U + U^\dagger)}
\Big \rangle_{U \in U(l)}.
\end{equation}

\subsection{Relationship to Schur polynomials}
Consider the set of all semi-standard tableaux of shape $\lambda$ and
content $N$, with each occurence of $j$ in the numbering therein
weighted $w_j$. The total weight
\begin{equation}\label{Sc}
s_\lambda(w_1,\dots,w_N) :=
\sum_{{\rm semi}\mbox{-}{\rm standard \: tableux} \atop
{\rm shape } \: \lambda, {\rm content} \: N}
w_1^{\# 1's} w_2^{\# 2's} \cdots w_N^{\# N's}
\end{equation}
is a symmetric polynomial in $w_1,\dots,w_N$ known as the Schur polynomial.
Because in a standard tableaux each number occurs exactly once, it follows
immediately that
\begin{equation}\label{tu.1}
[w_1 w_2 \cdots w_N] s_\lambda(w_1,\dots, w_N) = f_N^\lambda
\end{equation}
where $[w_1 w_2 \cdots w_N]$ denotes the coefficient of $w_1 w_2 \cdots w_N$.
This fact allows expressions such as (\ref{4.1}) involving $f_N^\mu$ to
be evaluated as special cases of Schur function identities. Explicitly, in
relation to (\ref{4.1}) use can be made of the Schur function identity
\begin{equation}\label{tu.2}
\sum_{\mu: \mu_1 \le l}
s_\mu(a_1,\dots,a_N) s_\mu(b_1,\dots,b_N) =
\Big \langle \prod_{j=1}^N \prod_{k=1}^l (1 + a_j e^{i \theta_k})
(1 + b_j e^{-i \theta_k}) \Big \rangle_{U(l)}.
\end{equation}
As to be revised in Section 4, this has direct relevance to the Johansson
model. Our present interest is that by extracting the coefficient of
$a_1 \cdots a_N b_1 \cdots b_N$ from both sides, making use of (\ref{tu.1})
on the left hand side, it follows that
$$
\sum_{\mu: \mu_1 \le l} (f_N^\mu)^2 = \Big \langle \Big | \sum_{k=1}^l 
e^{ i \theta_k} \Big |^{2N} \Big \rangle_{U(l)}
$$
in accordance with the evaluation of (\ref{4.1}) implied by the first equality
in (\ref{27.1}).
 
\subsection{A relationship to eigenvalue distributions}
It is of interest to note that the random matrix average over $U(l)$
in (\ref{27.2}) also arises in another probabilistic setting
\cite{Fo93c}. Thus consider the Laguerre unitary ensemble, specified by
the eigenvalue probability density function proportional to
$$
\prod_{l=1}^N \lambda_l^a e^{-\lambda_l} \prod_{1 \le j < k \le N}
(\lambda_k - \lambda_j)^2, \qquad \lambda_l > 0.
$$
For $a=n_1-n_2$, $n_1 \ge n_2$, it is realized by the eigenvalues of the
random matrix $X^\dagger X$, where $X$ is an $n_1 \times n_2$ complex
Gaussian matrix. Let $E_2^{\rm L}(s;a;N)$ denote the probability that there are
no eigenvalues in the interval $(0,s)$. Then it was shown in \cite{Fo93c}
that the scaled probability
\begin{equation}\label{27.3}
E_2^{\rm hard}(\lambda;a) := \lim_{N \to \infty} 
E_2^{\rm L}\Big ({\lambda \over N};a;N \Big ) 
\end{equation}
is given by the right hand side of (\ref{27.2}) with $l=a$ and
$\lambda \mapsto \sqrt{\lambda}$
(the superscript ``hard'' is used to denote the fact the eigenvalue
density is strictly zero to the left of $\lambda = 0$). This fact
has been used in \cite{BF03} to give a straightforward proof that
\begin{equation}\label{De}
\lim_{\lambda \to \infty} {\rm Pr} \Big ( {l^\symmU - 2 {\lambda}
\over \lambda^{1/3} } \le s \Big ) = F_{\rm GUE}(s).
\end{equation}
where $F_{\rm GUE}(s)$ is the scaled cumulative distribution of the largest
eigenvalue for large random Hermitian matrices with complex Gaussian
matrices, a celebrated result due originally to Baik, Deift and Johansson
\cite{BDJ98}, which gives as a corollary the scaled distribution of the
longest increasing subsequence of a random permutation.

\section{Symmetrizations of the Hammersley process}
\setcounter{equation}{0}
\subsection{Four symmetries of the square}
Baik and Rains \cite{BR01a} have formulated and analyzed four symmetrized
versions of the Hammersley process, in which the points are constrained
to have particular reflection symmetries of the square. In these
symmetrizations, the points are constrained to be invariant under
reflections about the diagonal $(0,0)$ to $(1,1)$;
diagonal $(0,1)$ to $(1,0)$; both these diagonals; and about the
centre point $(1/2,1/2)$. The four cases are denoted
${\symmO}$, ${\symmS}$, 
${\symmUU}$ and ${\symmu}$ respectively. The first three
of these cases can be further generalized to allow independent
Poisson rates for points forming on the diagonal(s). Let us
consider each case separately.
\subsection{The symmetry ${\symmO}$}
First we specify the Poisson process by which the points are added
to the diagonal and below the diagonal. We start with a time interval
$[0,z]$, which is broken up into $M$ smaller intervals of equal size,
the latter being labelled $j=1,\dots,M$. In each of these smaller
intervals, add one point below the diagonal (together with its image above
the diagonal) with probability $z^2(j-1/2)/M^2$ and a point on the
diagonal with probability $\alpha z/M$. The probability that
there are $n$ points in the square is then given by the coefficient of
$w^n$ in
$$
\prod_{j=1}^M \Big (1 - {\alpha \over M} + {\alpha w \over M} \Big )
\Big ( 1 - {(j-1/2) z^2 \over M^2} + {(j-1/2)z^2 w^2 \over M^2}
\Big ).
$$
It follows by taking the limit $M \to \infty$ in this expression that
the probability of their being exactly $n$ points after time $z$,
when the points below the diagonal are added with rate $z dz$ and
those below the diagonal with rate $\alpha dz$ is given by
\begin{equation}\label{2.dc1}
e^{-\alpha z} e^{-z^2/2} {z^n \over n!}
\sum_{m=0}^{[n/2]} \alpha^{n-2m} s_{n,m},
\qquad s_{n,m}:= \Big ( {n \atop 2m} \Big ) {(2m)! \over 2^m m!}.
\end{equation}
Furthermore, one sees that this probability conditioned so that there are
$n-2m$ points on the diagonal is equal to the $m$th term in this sum.

Let $l_{n,m}^\symmO$ denote the longest path length
in a realization of the $\symmO$ symmetrized Hammersley process consisting
of a total of $n$ points, $m$ of which are below the diagonal. It follows from
(\ref{2.dc1}) and the sentence below that if the points are chosen according
to the Poisson process specified above, then after time $z$
\begin{equation}\label{2.dc2}
{\rm Pr}(l^\symmO \le l) = e^{-\alpha z - z^2/2}
\sum_{n=0}^\infty {z^n \over n!} \sum_{m=0}^{[n/2]}
\alpha^{n-2m} s_{n,m} {\rm Pr}(l_{n,m}^\symmO \le l).
\end{equation}
To compute ${\rm Pr}(l_{n,m}^\symmO \le l)$ we first note that the sought
realizations of the $\symmO$ symmetrized Hammersley process are in
correspondence with $n \times n$ permutation matrices constrained to have 
$n - 2m$ non-zero entries on the diagonal, and which are furthermore symmetric
about the diagonal (recall that our convention is to count rows from the
bottom and so the diagonal runs from the bottom left to the top right).
It is straightforward to show that there are $s_{n,m}$ distinct such
permutation matrices.
Next we make a correspondence between such permutation matrices
and pairs of suitably constrained u/rh lattice paths, or equivalently
pairs of suitably constrained standard tableaux.

Now it is immediate from the rules of the PNG model that if the 
non-negative integer matrix $X=[x_{i,j}]_{i,j=1,\dots,n}$ maps to a pair
of u/rh paths $(P_1,P_2)$, then the transposed matrix $X^T=[x_{j,i}]_{i,j=
1,\dots,n}$ maps to a pair of u/rh paths $(P_2,P_1)$. Thus in the case that
$X = X^T$ one has $P_1=P_2$ so the mapping then is to a single family of
paths. 
(The example of Figure \ref{NLR.f1} exhibits this.) 
Furthermore, for symmetric non-negative integer matrices $X$ it
is a known  property of the RSK correspondence
that \cite{Kn70}
\begin{equation}\label{28.c}
\sum_{j=1}^n x_{j,j} = \sum_{j=1}^n (-1)^{j-1} \mu_j,
\end{equation}
a fact which can also be derived within the setting of the PNG model
\cite{FR02b}. Hence it follows that
\begin{eqnarray}\label{28.a1}
{\rm Pr}(l_{n,m}^\symmO \le l) & = & {1 \over s_{n,m}}
[\alpha^{n-2m}] \sum_{\mu:\mu_1 \le l}
\alpha^{\sum_{j=1}^n (-1)^{j-1} \mu_j} f_n^\mu \nonumber \\
& = & {1 \over s_{n,m}}
[\alpha^{n-2m}q_1 \cdots q_n ] \sum_{\mu:\mu_1 \le l}
\alpha^{\sum_{j=1}^n (-1)^{j-1} \mu_j}
s_\mu(q_1,\dots, q_n)
\end{eqnarray}
where the second equality follows from (\ref{tu.1}). Now  we know
from \cite{BR01a} that
\begin{equation}\label{29.5}
\sum_{\mu:\mu_1 \le l}
\alpha^{\sum_{j=1}^n (-1)^{j-1} \mu_j}
s_\mu(q_1,\dots, q_n) =
\Big \langle  \prod_{k=1}^l\Big ( (1 + \alpha e^{i \theta_k})
\prod_{j=1}^n(1 + q_j e^{i \theta_k})\Big )  \Big \rangle_{{\rm O}(l)}.
\end{equation}
In (\ref{29.5}) the average over the classical group O$(l)$ breaks into
two parts,
$$
\langle \: \cdot \: \rangle_{{\rm O}(l)} = {1 \over 2} \Big (
\langle \: \cdot \: \rangle_{{\rm O}^+(l)} +
\langle \: \cdot \: \rangle_{{\rm O}^-(l)} \Big )
$$
where $\langle \: \cdot \: \rangle_{{\rm O}^+(l)}$ 
denotes an average with respect
to the eigenvalue p.d.f.~for random matrices from the classical group
${\rm O}^+(l)$,
\begin{eqnarray}
&& {1 \over \pi^{l/2} 2^{l-1} (l/2)!}
\prod_{j=1}^{l/2}\delta(\theta_j - \theta_{l/2+j})
\prod_{1 \le j < k \le l/2} |e^{i \theta_j} - e^{i \theta_k}|^2
|1 - e^{i(\theta_j + \theta_k)}|^2, \qquad l \: {\rm even} \label{O.1}\\
&& {1 \over \pi^{(l-1)/2} 2^{l-1} ((l-1)/2)!}
 \delta(\theta_l) \prod_{j=1}^{(l-1)/2} 
\delta(\theta_j - \theta_{(l-1)/2+j})
|1- e^{i \theta_j}|^2
\nonumber \\
&& \qquad \times
\prod_{1 \le j < k \le (l-1)/2}|e^{i \theta_j} - e^{i \theta_k}|^2
|1 - e^{i(\theta_j + \theta_k)}|^2, \qquad l \: {\rm odd},
\nonumber \\
\end{eqnarray}
and $\langle \: \cdot \: \rangle_{{\rm O}^-(l)}$ denotes an average with respect
to the eigenvalue p.d.f.~for random matrices from the classical group
${\rm O}^-(l)$,
\begin{eqnarray}
&& {1 \over \pi^{l/2-1} 2^{l-2}  (l/2)!} \delta(\theta_{l-1})
\delta(\theta_l - \pi) \prod_{k=1}^{l/2 - 1} 
\delta(\theta_k - \theta_{l/2+k-1})
|1 - e^{2i \theta_k}|^2 \nonumber \\
&& \qquad \times
\prod_{1 \le j < k \le l/2-1} |e^{i \theta_j} - e^{i \theta_k}|^2
|1 - e^{i(\theta_j + \theta_k)}|^2, \qquad l \: {\rm even} \nonumber \\
\label{O.2}\\
&& {1 \over \pi^{(l-1)/2} 2^{l-1} ((l-1)/2)!}
\delta(\theta_l - \pi) \prod_{j=1}^{(l-1)/2} 
\delta(\theta_j - \theta_{(l-1)/2+j})
|1+ e^{i \theta_j}|^2 \nonumber \\
&& \qquad \times
\prod_{1 \le j < k \le (l-1)/2}|e^{i \theta_j} - e^{i \theta_k}|^2
|1 - e^{i(\theta_j + \theta_k)}|^2, \qquad l \: {\rm odd}.
\nonumber \\
\end{eqnarray}

Substituting (\ref{29.5}) in (\ref{28.a1}) we see that
\begin{equation}
{\rm Pr}(l_{n,m}^\symmO \le l) = 
{1 \over s_{n,m}} [\alpha^{n-2m}]
\Big \langle \prod_{k=1}^l(1 + \alpha e^{i \theta_k})
\Big ( \sum_{j=1}^l e^{i \theta_j} \Big )^n \Big \rangle_{{\rm O}(l)},
\end{equation}
and substituting this in (\ref{2.dc2}) then gives
\begin{equation}\label{3.20}
{\rm Pr}(l^\symmO \le l) =
e^{- \alpha z - z^2/2}
\Big \langle  \prod_{k=1}^l (1 + \alpha e^{i \theta_k})
e^{z \sum_{j=1}^l e^{i \theta_k}}  \Big \rangle_{{\rm O}(l)}
\end{equation}
as first obtained by Baik and Rains \cite{BR01a}.

As for the average (\ref{27.2}), the average (\ref{3.20}) in the case
$\alpha = 0$ also arises as a gap probability at the hard edge of a
matrix ensemble. Thus consider the Laguerre symplectic ensemble, specified
by the eigenvalue probability density function proportional to
$$
\prod_{l=1}^N \lambda_l^a e^{- \lambda_l} \prod_{1 \le j < k \le N}
|\lambda_k - \lambda_j|^4, \qquad \lambda_l > 0.
$$
For $a=2(n-N)+1$, $n \ge N$, this is realized by the distinct eigenvalues of
matrices of the form $X^\dagger X$ with $X$ an $n \times N$ real
quaternion Gaussian matrix (embedded as a complex matrix). Let
$E_4^{\rm L}(s;a;N)$ denote the probability that there are no eigenvalues
in the interval $(0,s)$. Then  we know from \cite{FW02,BF03} that for
$a \in \zz_{\ge 0}$
\begin{equation}\label{j.O}
E_4^{\rm hard}(\lambda;2a) :=
\lim_{N \to \infty} E_4^{\rm L}\Big ( {\lambda \over N}; 2a; N/2 \Big )
= e^{ - \lambda / 2} \Big \langle e^{\sqrt{\lambda} \sum_{j=1}^{2a}
e^{i \theta_j}} \Big \rangle_{{\rm O}(2a)},
\end{equation}
thus coinciding with (\ref{3.20}) in the case $\alpha = 0$ and $z^2=\lambda$,
$l=2a$.

\subsection{The symmetry ${\symmS}$}
Analogous to the case of the symmetry $\symmO$, let us suppose points are
added below the diagonal from $(0,1)$ to $(1,0)$ (to be referred to as the
anti-diagonal) with rate $z dz$. Let the mirror images of these points in
the anti-diagonal be also added, and let points on the anti-diagonal be added
with rate $\beta dz$. Then we know from (\ref{2.dc1}) and the 
following sentence that the probability there are exactly $n$ points after
time $z$, $n-2m$ of which are on the anti-diagonal, is equal to
$$
e^{- \beta z} e^{- z^2/2} {z^n \over n!} \beta^{n-2m} s_{n,m}.
$$

Now the permutation of $\{1,2,\dots,n\}$ corresponding to a realization of
this process is closely related to a realization of the symmetry
${\symmO}$. Thus let $P=(P(1),P(2),\dots,P(n))$ be a permutation with the
property that if $P(j) = k$ then $P(k) = j$ and thus correspond to the
symmetry $\symmO$. Then $\tilde{P} := (P(n), P(n-1), \dots, P(1))$
has the property that if $\tilde{P}(j) = k$ then $\tilde{P}(n+1-k) =
n+1 - j$ and so corresponds to the symmetry 
$\symmS$. Consequently the
maximum length of the increasing subsequences in the case of
$\symmS$, $l_{n,m}^\symmS$ say, is equal to the maximum length of the
decreasing subsequences for $\symmO$. 
Furthermore, it follows from a theorem
of Greene \cite{Gr74} (see also \cite{Sa00}) relating row lengths of the
standard tableau $\kappa$ corresponding to $P$, to increasing
subsequences, and column lengths to decreasing subsequences, that the
conjugate tableaux $\kappa'$ obtained by interchanging the row and column
lengths in $\kappa$, corresponds to $P^R$. Also, it follows from
(\ref{28.c}) that
\begin{equation}\label{1.6.0}
\#(1\mbox{'}{\rm s \: on \: the \: anti}\mbox{-}{\rm diagonal}) =
\sum_{j=1}^n (-1)^{j-1} \kappa_j
\end{equation}
Setting $\kappa' = \mu$, these facts together imply
\begin{eqnarray}\label{1.6.1}
{\rm Pr}(l_{n,m}^\symmS \le l) & = & {1 \over s_{n,m}} [\beta^{n-2m}]
\sum_{\mu: \mu_1 \le l} \beta^{\sum_{j=1}^n (-1)^{j-1} \mu_j'}
f_n^\mu \nonumber \\
& = &
{1 \over s_{n,m}} [\beta^{n-2m}q_1 \cdots q_n]
\sum_{\mu: \mu_1 \le l} \beta^{\sum_{j=1}^n (-1)^{j-1} \mu_j'}
s_\mu(q_1,\dots, q_n)
\end{eqnarray}
(cf.~(\ref{28.a1})). In \cite{BR01a} it was shown that
$$
\sum_{\mu : \mu_1 \le 2l}  \beta^{\sum_{j=1}^n (-1)^{j-1} \mu_j'}
s_\mu(q_1,\dots,q_n) = \Big \langle
\prod_{k=1}^{2l} \Big ( {1 \over 1 - \beta e^{-i \theta_k}}
\prod_{j=1}^n (1 + q_j e^{ i \theta_k}) \Big ) \Big \rangle_{{\rm Sp}(l)}
$$
and
$$
\sum_{\mu : \mu_1 \le 2l+1}  \beta^{\sum_{j=1}^n (-1)^{j-1} \mu_j'}
s_\mu(q_1,\dots,q_n) = \prod_{j=1}^n (1 + \beta q_j)
\Big \langle
\prod_{k=1}^{2l} 
\prod_{j=1}^n 
(1 + q_j e^{ i \theta_k})  \Big \rangle_{{\rm Sp}(l)}
$$
where $\langle \: \rangle_{{\rm Sp}(l)}$ denotes an average with respect to
the eigenvalue p.d.f.~for random matrices from the classical group
Sp$(l)$,
$$
{1 \over (2 \pi)^l} {1 \over 2^l l!} \prod_{k=1}^l 
\delta(\theta_k - \theta_{l+k})
|e^{i \theta_k} -
e^{-i \theta_k} |^2
\prod_{1 \le j < k \le l} | e^{i \theta_j} - e^{i \theta_k}|^2
|1 - e^{i(\theta_j + \theta_k)} |^2.
$$
It follows immediately that
\begin{eqnarray*}
{\rm Pr}(l_{n,m}^\symmS \le 2l) & = & {1 \over s_{n,m}} [\beta^{n-2m}]
\Big \langle
\prod_{k=1}^{2l}  {1 \over 1 - \beta e^{-i \theta_k}}
\Big ( \sum_{j=1}^{2l} e^{i \theta_j} \Big )^n \Big \rangle_{{\rm Sp}(l)}
\\
{\rm Pr}(l_{n,m}^\symmS \le 2l+1) & = & {1 \over s_{n,m}} [\beta^{n-2m}]
\Big \langle
\Big ( \beta + \sum_{j=1}^{2l} 
e^{i \theta_j} \Big )^n \Big \rangle_{{\rm Sp}(l)}.
\end{eqnarray*}
Thus with
\begin{equation}
{\rm Pr}(l^\symmS \le l) := e^{-\beta z - z^2/2}
\sum_{n=0}^\infty { z^n \over n!} \sum_{m=0}^n \beta^{n-2m} s_{n,m}
{\rm Pr}(l_{n,m}^\symmS \le l)
\end{equation}
we have \cite{BR01a}
\begin{eqnarray}\label{4.24}
{\rm Pr}(l^\symmS \le 2l) & = & e^{-\beta z - z^2/2} 
\Big \langle e^{z \sum_{j=1}^{2l} e^{i \theta_j}  }
\prod_{k=1}^{2l}  {1 \over 1 - \beta e^{-i \theta_k}}
\Big \rangle_{{\rm Sp}(l)} \nonumber \\
{\rm Pr}(l^\symmS \le 2l +1) & = & e^{- z^2/2}
\Big \langle e^{z \sum_{j=1}^{2l} e^{i \theta_j}  }
\Big \rangle_{{\rm Sp}(l)}.
\end{eqnarray}

There is an analogue of (\ref{j.O}) for the second average in
(\ref{4.24}). For this consider the Laguerre orthogonal ensemble,
specified by the eigenvalue probability density function proportional to
$$
\prod_{l=1}^N \lambda_l^a e^{- \lambda_l / 2} \prod_{1 \le j < k \le N}
| \lambda_k - \lambda_j|, \qquad \lambda_l > 0.
$$
For $a=(n-N-1)/2$, $n \ge N$, this is realized by the distinct eigenvalues of
matrices of the form $X^T X$ with $X$ an $n \times N$ real standard
Gaussian matrix. Let
$E_1^{\rm L}(s;a;N)$ denote the probability that there are no eigenvalues
in the interval $(0,s)$. Then  we know from \cite{FW02,BF03} that for
$a \in \zz_{\ge 0}$
\begin{equation}\label{j.Oa}
E_1^{\rm hard}(\lambda;a) :=
\lim_{N \to \infty} E_1^{\rm L}\Big ( {\lambda \over N}; a; N \Big )
= e^{ - \lambda / 2} \Big \langle e^{\sqrt{\lambda} \sum_{j=1}^{2a}
e^{i \theta_j}} \Big \rangle_{{\rm Sp}(a)},
\end{equation}
thus coinciding with (\ref{4.24}) after setting $\lambda = z^2$, $a=l$.

\subsection{The symmetry $\symmu$}
For this symmetry only the points below both the diagonal and anti-diagonal
are independent. Let us suppose points are added to this region with
rate $2z dz$. Suppose too that points are added to the bottom half of
the diagonal with rate $\alpha dz$, and to the bottom half of the
anti-diagonal with rate $\beta dz$. The images of all these points must also
be added. The probability of there being exactly $2n$ points after time $z$,
$2m_+$ of which are on the diagonal, and $2m_-$ of which are on the
anti-diagonal, is then equal to
$$
{z^n \over n!} e^{ - z^2 - \alpha z - \beta  z} \alpha^{m_+}
\beta^{m_-} t_{n, m_+, m_-}
$$
where 
$$
t_{n, m_+, m_-} = {n! \over m_+! m_-! ((n-m_+ - m_-)/2)!}
$$
and it is required $n-m_+-m_-$ be even.

The permutation matrix $P=[x_{i,j}]_{i,j=1,\dots,2n}$ corresponding to a
realization of this process has the two symmetries $P = P^T =
[x_{j,i}]_{i,j=1,\dots,2n}$ and $P=P^R=[x_{2n+1-j, 2n+1-i}]_{i,j=1,\dots,2n}$.
Now whereas a permutation matrix with the symmetry $P=P^T$ maps under the RSK
correspondence to a pair of identical standard tableaux $(T_1, T_1)$ say, a
permutation matrix with the symmetry $P=P^R$ maps to a pair of standard
tableaux $(T_2^R, T_2)$ where $T_2^R$ denotes the Sch\"utzenberger dual of
$T_2$ (see e.g.~\cite{Sa00}). We note too that the number of permutation
matrices of $\{1,2,\dots,2n\}$ with the symmetry $P=P^T=P^R$ such that 
$m_+$ ($m_-$) members of $\{1,2,\dots,n\}$ have the property that
$P(i) = i$ ($P(i) = 2n+1 - i$) is equal to $t_{n, m_+, m_-}$.
Hence with $l^\symmu_{n, m_+, m_-}$ denoting the longest path length in a
realization of the Hammersley process with symmetry $\symmu$, we
have
\begin{equation}
{\rm Pr}(l^\symmu_{n, m_+, m_-} \le l) = {[\alpha^{m_+} \beta^{m_-}] \over
t_{n,m_+,m_-}} \sum_{\mu: \mu_1 \le l}
\alpha^{\sum_{j=1}^{2n} (-1)^{j-1} \mu_j} \beta^{\sum_{j=1}^{2n}
(-1)^{j-1} \mu_j'} \tilde{f}_{2n}^\mu
\end{equation}
where
$$
\tilde{f}_{2n}^\mu = \# ({\rm self \: dual \: standard \: tableaux, \: shape}
\: \mu, \: {\rm content} \: 2n).
$$

Analogous to (\ref{Sc}), let us define
\begin{equation}\label{Sc1}
\tilde{s}_\lambda^{\rm s.d.}(q_1,\dots,q_n) = \sum\nolimits^*
q_1^{\# 1's} \cdots q_n^{\# n's}
\end{equation}
where the asterisk denotes that the sum is over all self-dual
semi-standard tableaux of shape $\lambda$ and content $2n$.
(For self dual tableaux of content $2n$, $\#j\mbox{'}s 
= \# (2n+1-j)\mbox{'}s$ so
there are $n$ rather than $2n$ independent variables in the weightings.)
From the definitions
$$
\tilde{f}_{2n}^\mu = [q_1 \cdots q_n] 
\tilde{s}_\lambda^{\rm s.d.}(q_1,\dots,q_n),
$$
so we can write
\begin{eqnarray*}
{\rm Pr}(l^\symmu_{n,m_+,m_-} \le l) & = & {1 \over t_{n,m_+,m_-}}
[\alpha^{m_+} \beta^{m_-} q_1 \cdots q_n]
\sum_{\mu: \mu_1 \le l} \alpha^{\sum_{j=1}^{2n} (-1)^{j-1} \mu_j}
\nonumber \\
&& \quad \times
\beta^{\sum_{j=1}^{2n} (-1)^{j-1} \mu_j'}
\tilde{s}_\lambda^{\rm s.d.}(q_1,\dots,q_n).
\end{eqnarray*}
Baik and Rains \cite{BR01a} have provided the evaluation
$$
\sum_{\mu: \mu_1 \le 2l} \alpha^{\sum_{j=1}^{2n}
(-1)^{j-1} \mu_j} \beta^{\sum_{j=1}^{2n} (-1)^{j-1} \mu_j'}
\tilde{s}_\lambda^{\rm s.d.}(q_1,\dots,q_n)
=
\Big \langle \prod_{j=1}^l {1 + \alpha e^{i \theta_j} \over 1 - \beta
e^{i \theta_j} } \prod_{k=1}^n |1 + q_k e^{ i \theta_j} |^2
\Big \rangle_{U(l)}.
$$
Hence
$$
{\rm Pr}(l^\symmu_{n,m_+,m_-} \le 2l) = [\alpha^{m_+} \beta^{m_-}]
\Big \langle \prod_{j=1}^l {1 + \alpha e^{i \theta_j} \over 1 - \beta
e^{i \theta_j} } \Big ( \sum_{k=1}^l 2 \cos \theta_k \Big )^n
\Big \rangle_{U(l)}
$$
and consequently with
$$
{\rm Pr}(l^\symmu \le 2l) := e^{- z^2 - \alpha z - \beta z}
\sum_{n=0}^\infty {z^n \over n!} \sum_{m_+, m_- \ge 0}
\alpha^{m_+} \beta^{m_-} t_{n,m_+,m_-}
{\rm Pr}(l^\symmu_{n,m_+,m_-} \le 2l)
$$
we obtain the result \cite{BR01a}
\begin{equation}
{\rm Pr}(l^\symmu \le 2l) = e^{- z^2 - \alpha z - \beta z}
\Big \langle \prod_{j=1}^l {1 + \alpha e^{i \theta_j} \over 1 - \beta
e^{i \theta_j} } e^{2 z \sum_{k=1}^l \cos \theta_k}
\Big \rangle_{U(l)}.
\end{equation}
Baik and Rains \cite{BR01a} have also provided the evaluation
\begin{eqnarray*}
&&\sum_{\mu: \mu_1 \le 2l+1} \alpha^{\sum_{j=1}^{2n} (-1)^{j-1} \mu_j}
\beta^{\sum_{j=1}^{2n} (-1)^{j-1} \mu_j'}
\tilde{s}^{\rm s.d.}_\mu(q_1,\dots, q_n)  \\
&& \qquad = \prod_{k=1}^n(1 + \beta q_k)
\Big \langle \prod_{j=1}^l 
(1 + \alpha e^{ i \theta_j})
\prod_{k=1}^n|1 + q_k e^{ i \theta_j} |^2 \Big \rangle_{U(l)}.
\end{eqnarray*}
We readily deduce from this that \cite{BR01a}
\begin{equation}
{\rm Pr}(l^\symmu \le 2l+1) = {\rm Pr}(l^\symmu \le 2l) \Big |_{\beta = 0}.
\end{equation}

\subsection{The symmetry $\symmUU$}
In the unit square suppose points are marked in the region $y < 1/2$ with
Poisson rate $4 z dz$. For each point $(x',y')$ so marked, also mark the
image $(1-x',1-y')$, which corresponds to a reflection about the centre
of the square. A realization of this process with $2n$ points corresponds
to a permutation matrix with the symmetry $X = (X^R)^T$, or equivalently
to a permutation of $\{1,2,\dots,2n\}$ with the property that if
$P(i) = j$, then $P(2n+1-i) = 2n+1-j$. Note that there are
$(2n)!! = 2^n n!$ permutations of this type.

Now we know \cite{Sa00} that if a permutation matrix $P$ maps to a pair
of standard tableaux $(T_1,T_2)$ under the Robinson-Schensted mapping,
then $P^R$ maps to $(T_2^R, T_1^R)$ while $P^T$ maps to $(T_2,T_1)$. Hence
a permutation matrix with the symmetry $X=(X^R)^T$ maps to a pair of
standard tableaux $(T_1,T_2)$ constrained so that
$$
T_1 = T_1^R, \qquad T_2 = T_2^R.
$$
Consequently
with $l_{2n}^\symmUU$ denoting the longest path in a realization of
$2n$ points,
\begin{eqnarray*}
{\rm Pr}(l_{2n}^\symmUU \le l) & = & {1 \over 2^n n!} \sum_{\mu: \mu_1 \le l}
(\tilde{f}_{2n}^\mu )^2 \\
& = & {1 \over 2^n n!} [a_1 \cdots a_n b_1 \cdots b_n]
\sum_{\mu: \mu_1 \le l}
\tilde{s}_\mu(a_1,\dots,a_n) \tilde{s}_\mu(b_1,\dots,b_n).
\end{eqnarray*}

Baik and Rains \cite{BR01a} have derived the results
\begin{eqnarray*}
&& \sum_{\mu: \mu_1 \le 2l}
\tilde{s}_\mu(a_1,\dots,a_n) \tilde{s}_\mu(b_1,\dots,b_n) \\
&& \qquad =
\Big ( \Big \langle \prod_{j=1}^n \prod_{k=1}^l (1 + a_j e^{i \theta_k})
(1 + b_j e^{-i \theta_k}) \Big \rangle_{U(l)} \Big )^2 
 = \Big \langle \prod_{j=1}^n \prod_{k=1}^{2l}(1 + a_j e^{i \theta_k})
(1 + b_j e^{-i \theta_k}) \Big \rangle_{U(l)\oplus U(l)}.
\end{eqnarray*}
\begin{eqnarray*}
&& \sum_{\mu: \mu_1 \le 2l+1}
\tilde{s}_\mu(a_1,\dots,a_n) \tilde{s}_\mu(b_1,\dots,b_n) \\
&& \qquad =
\Big \langle \prod_{j=1}^n \prod_{k=1}^{l+1} (1 + a_j e^{i \theta_k})
(1 + b_j e^{-i \theta_k}) \Big \rangle_{U(l+1)}
\Big \langle \prod_{j=1}^n \prod_{k=1}^l (1 + a_j e^{i \theta_k})
(1 + b_j e^{-i \theta_k}) \Big \rangle_{U(l)} \\
&& \qquad =
\Big \langle \prod_{j=1}^n \prod_{k=1}^{2l+1}(1 + a_j e^{i \theta_k})
(1 + b_j e^{-i \theta_k}) \Big \rangle_{U(l+1)\oplus U(l)}.
\end{eqnarray*}
Consequently
\begin{eqnarray}
{\rm Pr}(l_{2n}^\symmUU \le 2l) & = & {1 \over 2^n n!}
\Big \langle \Big | \sum_{k=1}^{2l} e^{ i \theta_k} \Big |^{2n}
\Big \rangle_{U(l)\oplus U(l)} \nonumber \\
{\rm Pr}(l_{2n}^\symmUU \le 2l+1) & = & {1 \over 2^n n!}
\Big \langle \Big | \sum_{k=1}^{2l+1} e^{ i \theta_k} \Big |^{2n}
\Big \rangle_{U(l+1)\oplus U(l)}
\end{eqnarray}
and thus with
$$
{\rm Pr}(l^\symmUU \le l) := e^{- 2 z^2} \sum_{n=0}^\infty {2^n z^{2n} \over
n!} {\rm Pr}(l^\symmUU_{2n} \le l)
$$
we see that \cite{BR01a}
\begin{eqnarray}
{\rm Pr}(l^\symmUU \le 2l) & = & \Big ( {\rm Pr}(l^\symmU \le l)
\Big |_{\lambda \mapsto z} \Big )^2 \nonumber \\
{\rm Pr}(l^\symmU \le 2l) & = &
{\rm Pr}(l^\symmU \le l+1) \Big |_{\lambda \mapsto z}
{\rm Pr}(l^\symmU \le l)
\Big |_{\lambda \mapsto z}
\end{eqnarray}
where ${\rm Pr}(l^\symmU \le l)$ is given by (\ref{27.2}).

\section{The Hammersley process with sources on the boundary}
\setcounter{equation}{0}
The Hammersley process which relates to the scaling function for
one-dimensional stationary KPZ growth is the original model (Poisson
points in a square), generalized to  allow independent Poisson rates for
points forming on the boundaries $y=0$ (with intensity $\alpha_-$) and
$x=0$ (with intensity $\alpha_+$). 
In the PNG model picture, these boundary points correspond to growth
from the boundary of the expanding droplet, with rate $\alpha_+ dt$ on the
left boundary, and rate $\alpha_- dt$ on the right boundary. Such
boundary growths can be realized by an initial condition of mean slope
$-\alpha_-$ $(\alpha_+)$ for $x<0$ $(x > 0)$, created by a staircase
structure (downward sloping for $x < 0$, upward sloping for $x>0$),
with vertical increments of one unit with intensity $\alpha_-$
($\alpha_+$).
A formula for the cumulative probability of
the longest increasing subsequence length of this model has been obtained
by Baik and Rains \cite{BR00}. This was obtained as a limiting case of an
inhomogeneous lattice generalization of the Hammersley process due to
Johansson \cite{Jo99a}. 

In the Johansson model, on each site $(i,j)$, $0 \le i,j \le n$ of a
$(n+1) \times (n+1)$ square grid there is a non-negative integer
geometric random variable $x_{i,j}$ with parameter $a_i b_j$ so that
\begin{equation}\label{10.6.1}
{\rm Pr}(x_{i,j}=k) = (1 - a_i b_j) (a_i b_j)^k.
\end{equation}
A type of directed last passage percolation is to form an u/rh lattice path
from the site $(0,0)$ to the site $(n,n)$ such that it maximizes the sum total
of the random variables associated with the sites. Of interest then is the 
value of this maximized sum,
$$
L(n,n) := {\rm max} \sum_{(0,0) {\rm u/rh} (n,n)} x_{i,j}.
$$
It is not hard to show that in the RSK correspondence mapping the
non-negative integer matrix $X = [x_{i,j}]_{i,j=0,\dots,n}$ to a pair
of semi-standard tableaux of the same shape and content $n+1$, the length
of the first row of the tableaux is equal to $L(n,n)$ (one approach
is to show that $L(n,n)$ and the height at the origin in the PNG model 
picture of the RSK correspondence satisfy the same recurrence). Further, one
has that with entries of the matrix chosen according to (\ref{10.6.1}), the
RSK correspondence maps to pairs of weighted semi-standard tableaux,
weights $\{a_i\}$, $\{b_j\}$ repsectively. Consequently
$$
{\rm Pr}(L(n,n) \le l) = \prod_{j,k=0}^n(1 - a_j b_k)
\sum_{\mu: \mu_1 \le l} s_\mu(a_0,\dots, a_n) s_\mu(b_0,\dots, b_n)
$$
and thus according to (\ref{tu.2})
\begin{equation}\label{10.6.4}
{\rm Pr}(L(n,n) \le l) = \prod_{j,k=0}^n(1 - a_j b_k)
\Big \langle \prod_{j=0}^n \prod_{k=1}^l(1 + a_j e^{i \theta_k})(1 +
b_j e^{-i \theta_k}) \Big \rangle_{U(l)}.
\end{equation}

Suppose we set 
\begin{equation}\label{10.6.5}
a_i = b_i = t/n \qquad (i=1,\dots,n)
\end{equation}
and take the limit $n \to \infty$. Then the probability that there are
$k$ points in the region $1 \le i,j \le n$ has the large $n$ behaviour
\begin{equation}\label{10.6.5a}
\Big ( {n^2 \atop k} \Big )
\Big ( \Big ( 1 - {t^2 \over n^2} \Big ) {t^2 \over n^2} \Big )^k
\Big ( 1 - {t^2 \over n^2} \Big )^{n^2 - k} \to
{e^{-t^2} t^{2k} \over k!}
\end{equation}
where the first factor corresponds to the number of different ways of choosing
$k$ sites from the grid of $n^2$ sites, 
the second factor is the probability that those $k$
sites are occupied, while the final factor is the probability that the
remaining sites are empty. Hence we reclaim the setting of the original
Hammersley model with Poisson parameter $t^2$. Indeed with $a_0=b_0=0$
we see that by substituting (\ref{10.6.5}) in (\ref{10.6.4}) and taking the
limit $n \to \infty$ (\ref{27.2}) results with $\lambda = t$.

Instead of setting $a_0$ and $b_0$ to zero, suppose we choose 
$a_0 = \alpha_+$, $b_0 = \alpha_1$. Then the argument leading to
(\ref{10.6.5a}) shows that along $x>0$ $(y > 0)$ we obtain a Poisson
process of intensity $\alpha_+ t$ ($\alpha_- t$). At the origin there remains
a non-negative integer variable chosen according to the
geometric distribution with parameter $\alpha_+ \alpha_-$,
$g(\alpha_+ \alpha_-)$. Let us denote by 
$L^+(t,\alpha_+,\alpha_-)$ the length of the longest
up/right path in this process, and let us denote by 
$L(t,\alpha_+,\alpha_-)$ the same quantity
but with the geometric random variable at the origin removed. Clearly
\begin{equation}\label{11.6.1}
L^+(t,\alpha_+,\alpha_-) = 
L(t,\alpha_+,\alpha_-) + \chi, \qquad \chi \in g(\alpha_+ \alpha_-),
\end{equation}
while the appropriate limit of (\ref{10.6.4}) gives
\begin{equation}\label{11.6.pa}
{\rm Pr}(L^+(t,\alpha_+,\alpha_-) \le l) = (1 - \alpha_+ \alpha_-)
e^{-(\alpha_+ + \alpha_-)t - t^2} \tilde{D}_l
\end{equation}
where
\begin{equation}\label{11.6.p}
\tilde{D}_l = \Big \langle \prod_{j=1}^l (1 + \alpha_+ e^{i \theta_j})
(1 + \alpha_- e^{-i \theta_j}) e^{2 t \sum_{j=1}^l \cos \theta_j}
\Big \rangle_{U(l)}.
\end{equation}
To now obtain a formula for ${\rm Pr}(L(t) \le l)$, introduce the generating
functions
$$
Q(x) = \sum_{l=0}^\infty {\rm Pr}(L(t) \le l) x^l, \qquad
Q^+(x) = \sum_{l=0}^\infty {\rm Pr}(L^+(t) \le l) x^l.
$$
Then using (\ref{11.6.1}) we see that
\begin{eqnarray*}
Q^+(x) & = & \sum_{l=0}^\infty x^l \sum_{k=0}^l {\rm Pr}(L(t) \le l - k)
{\rm Pr}(\chi = k) \\
& = & (1 - \alpha_+  \alpha_-) \sum_{l=0}^\infty x^l \sum_{k=0}^l
{\rm Pr}(L(t) \le l - k)
(\alpha_+ \alpha_-)^k \: = \: {1 - \alpha_+ \alpha_- \over 1 - x 
\alpha_+  \alpha_-} Q(x)
\end{eqnarray*}
where the final equality follows by writing $x^l = x^{l-k} x^k$ and summing 
independently over $l-k$ and $k$. Multiplying both sides of this identity
by $1 - x \alpha_+ \alpha_-$ and equating like powers of $x$ gives
\cite{BR00}
\begin{equation}\label{4.41}
{\rm Pr}(L(t,\alpha_+,\alpha_-) \le l) = 
e^{-(\alpha_+ + \alpha_-)t - t^2} (\tilde{D}_l - \alpha_+ \alpha_-
\tilde{D}_{l-1}).
\end{equation}

To proceed further, $\tilde{D}_l$ is expressed in terms of
\begin{equation}\label{11.d3}
D_l = \tilde{D}_l \Big |_{\alpha_+ = \alpha_- = 0} =
\Big \langle e^{2t \sum_{j=1}^l \cos \theta_j} \Big \rangle_{U(l)},
\end{equation}
and monic orthogonal polynomials $\{ \pi_j(e^{i \theta}) \}_{j=0,1,\dots}$
with respect to the weight $e^{2t \cos \theta}$,
\begin{equation}\label{11.d4}
{1 \over 2 \pi} \int_{-\pi}^\pi \pi_j(e^{i \theta}) \overline{
\pi_k(e^{i \theta})} e^{2t \cos \theta} \, d \theta = 
{1 \over
\kappa_j^2} \delta_{j,k}.
\end{equation}
For this purpose, let
\begin{equation}\label{6.40'}
\pi_n^*(z) := z^n \pi_n(z^{-1}).
\end{equation}
Then with $\{ \pi_j(e^{i \theta}) \}_{j=0,1,\dots}$ the monic 
orthogonal polynomials corresponding to the general real weight $w(\theta)$
replacing $e^{2t \cos \theta}$ in (\ref{11.d4}), we have the general formula
\cite{Sz75}
\begin{eqnarray}\label{6.41}
{ \Big \langle \prod_{j=1}^n w(\theta_j) (e^{i \theta_j} - x)
(e^{-i \theta_j} - y) \Big \rangle_{U(n)} \over
\Big \langle \prod_{j=1}^n w(\theta_j) \Big \rangle_{U(n)} } & = &
{\pi_{n+1}^*(x) \pi_{n+1}^*(y) - \pi_{n+1}(x) \pi_{n+1}(y) \over 1 - xy}
\nonumber \\
& = & {\pi_{n}^*(x) \pi_{n}^*(y) - xy \pi_n(x) \pi_n(y) \over 1 - xy}
\nonumber
\\ & = &
{1 \over \kappa_n^2} \sum_{k=0}^{n} \kappa_k^2 \pi_k(x) \pi_k(y).
\end{eqnarray}
Application of the second of these equalities shows
$$
\tilde{D}_l = {\pi^*_l(-\alpha_+) \pi^*_l(-\alpha_-) - \alpha_+
\alpha_- \pi_l(-\alpha_+) \pi_l(-\alpha_-) \over
1 - \alpha_+ \alpha_-} D_l
$$
and in particular, applying l'H\^opitals rule, we see that
\begin{equation}\label{D.D}
\tilde{D}_l \Big |_{\alpha_+=1/\alpha_- = \alpha} =
\Big \{ (1 - l) \pi_l(-\alpha) \pi_l(-\alpha^{-1}) - \alpha
\pi_l'(-\alpha) \pi_l(-\alpha^{-1}) - \alpha^{-1}
\pi_l(-\alpha) \pi_l'(-\alpha^{-1}) \Big \} D_l.
\end{equation}

The scaling behaviour of $D_l = D_l(t)$ as defined by (\ref{11.d3}) is the 
result (\ref{De}) of Baik, Deift and Johansson. Thus
we have
\begin{equation}\label{4.47}
\lim_{t \to \infty} e^{-t^2} D_{[2t + t^{1/3} s]}(t) = F_{\rm GUE}(s).
\end{equation}
We note that $F_{\rm GUE}(s)$
has the exact evaluation \cite{TW94a}
\begin{equation}\label{4.47a}
F_{\rm GUE}(s) = \exp \Big \{ - \int_s^\infty (t - s) q^2(t) \, dt
\Big \}
\end{equation}
where $q(t)$ is the solution of the non-linear equation 
\begin{equation}\label{11.6}
q'' = tq + 2 q^3
\end{equation}
(a special case of the Painlev\'e II equation) subject to the boundary
condition
\begin{equation}\label{11.7}
q(t) \sim - {\rm Ai}(t) \qquad {\rm as} \quad t \to \infty
\end{equation}
where ${\rm Ai}(t)$ denotes the Airy function.

In addition to setting
\begin{equation}\label{ls}
l =  [2t + t^{1/3} s],
\end{equation}
to obtain critical scaling behaviour, the parameter $\alpha$ must
be related to $t$ by \cite{BR00}
\begin{equation}\label{ay}
\alpha = 1 - y/ t^{1/3}
\end{equation}
where $y$ is fixed. To specify the corresponding
scaled form of the orthogonal polynomials
in (\ref{D.D}), one establishes equations for their variation with respect
to the scaled variables $s$ and $y$
\cite{BR00,Ba01,PS02}. In relation to the former, one recalls
that in general monic orthogonal polynomials on the unit circle satisfy the
coupled recurrences \cite{Sz75}
\begin{eqnarray}\label{13.0}
\pi_{n+1}(z) & = & z \pi_n(z) + r_{n+1} \pi_n^*(z) \nonumber \\
\pi_{n+1}^*(z) & = & r_{n+1} z \pi_n(z) +  \pi_n^*(z)
\end{eqnarray}
where $r_n = \pi_n(0)$. For the weight $w(z) = e^{2t \cos \theta}$ the
scaled form of $r_l = r_l(t)$ with the substitution (\ref{ls}) can be 
determined from a difference equation --- a form of the discrete Painlev\'e
II equation --- satisfied by $\{r_n\}$ \cite{PS90a,Hi96a,TW99a,IW01,Ba01,Bo01a,
AV02,FW03}.

\begin{prop}
With $r_n := \pi_n(0)$, the sequence $\{r_n \}$ for the polynomials with
orthogonality (\ref{11.d4}) satisfies a form of the discrete Painlev\'e
II equation
\begin{equation}\label{rt}
- {n \over t} {r_n \over 1 - r_n^2} = r_{n+1} + r_{n-1}
\end{equation}
subject to the initial conditions
\begin{equation}\label{rt.a}
r_0 = 1, \qquad r_1 = - {I_1(t) \over I_0(t)},
\end{equation}
where $I_\nu(t)$ denotes the Bessel function of pure imaginary argument.
\end{prop}

{\rm Proof.} \quad Following  \cite{IW01}, let us show how (\ref{rt})
can be derived using simple properties of the weight
\begin{equation}\label{rt.1}
w(z) = e^{t(z+ 1/z)},
\end{equation}
together with a general formula from the theory of orthogonal polynomials
on the unit circle. Regarding the latter, set $\phi_j(z) := \kappa_j 
\pi_j(z)$ so that according to (\ref{11.d4}) $\{\phi_j(z)\}$ defines
an orthonormal set of polynomials on the unit circle, and
introduce the coefficient $l_n$ by
\begin{equation}\label{rt.2}
\phi_n(z) = \kappa_n z^n + l_n z^{n-1} + \cdots + \phi_n(0).
\end{equation}
Then it is generally true that \cite{Sz75}
\begin{equation}\label{j.0}
{l_n \over \kappa_n} = \sum_{j=0}^{n-1} r_{j+1} \bar{r}_j, \qquad
r_j := {\phi_j(0) \over \kappa_j}.
\end{equation}
Specific to the weight (\ref{rt.1}) consider
$$
J :=  \int_{\cal C} z^2
\Big ( {d \over dz} w(z) \Big ) \phi_n(z) 
\overline{\phi_{n+1}(z)} \, {dz \over 2 \pi i z},
$$
where ${\cal C}$ is a simple closed contour encircling the origin.
Noting from (\ref{rt.1}) that
$$
{d \over dz} w(z) = t \Big ( 1 - {1 \over z^2} \Big ) w(z)
$$
we see from integration by parts, the structure (\ref{rt.2}), and the
orthonormality of $\{\phi_j(z)\}$ that
\begin{equation}\label{j.1}
J = - (n+1){ \kappa_n \over \kappa_{n+1}} + (n+1) {\kappa_{n+1} \over
\kappa_n}.
\end{equation}
On the other hand, direct evaluation of $J$ using (\ref{rt.2})
and the orthonormality of $\{\phi_j(z)\}$ shows
\begin{equation}\label{j.2}
J = t \Big (
{l_n \over \kappa_{n+1}} - {l_{n+2} \over \kappa_{n+2}}
{\kappa_n \over \kappa_{n+1}} \Big ).
\end{equation}
Equating (\ref{j.1}) and (\ref{j.2}) and eliminating $l_n$ using (\ref{j.0})
(with $\overline{r_j} = r_j$) we arrive at (\ref{rt}). \hfill $\square$

\medskip
Following \cite{PS02},
with
$$
R_n(t) := (-1)^{n-1} r_n(t)
$$
and making the ansatz
\begin{equation}\label{j.3}
R_{[2t + t^{1/3} s]}(t) \: \sim \: t^{-1/3} u(s), \qquad t \to \infty,
\end{equation}
we see that formally the difference equation (\ref{rt}) becomes the
differential equation
\begin{equation}\label{j.4}
{d^2 u \over d s^2} = su + 2 u^3.
\end{equation}
Further, for (\ref{j.3}) to be compatible with the first of the initial
conditions (\ref{rt.a}), one must have
\begin{equation}\label{13.2}
u(s) \mathop{\sim}\limits_{s \to - \infty} - \sqrt{-s/2}.
\end{equation}
Now (\ref{j.4}) is the same particular Painlev\'e II equation as (\ref{11.6}).
In fact it is a celebrated result \cite{HM80} in the theory of the Painlev\'e
II equation that (\ref{11.6}) has a unique solution with the
asymptotic $s \to - \infty$ behaviour (\ref{13.2}), and the asymptotic
$s \to \infty$ behaviour $u(s) \sim - {\rm Ai}(s)$ as in
(\ref{11.7}). Thus we conclude
\begin{equation}\label{13.3}
R_{[2t + t^{1/3} s]}(t) \: \sim \: t^{-1/3} q(s), \qquad t \to \infty.
\end{equation}

According to (\ref{D.D}), our interest is in $\pi_n(-\alpha)$ and
$\pi_n^*(-\alpha)$, so we should set $z=\alpha$ in (\ref{13.0}).
Introducing
\begin{equation}\label{sle1}
P_n(\alpha) = e^{-t \alpha} \pi_n^*(-\alpha), \qquad
Q_n(\alpha) = - e^{-t \alpha} (-1)^n \pi_n(-\alpha),
\end{equation}
one sees that (\ref{13.0}) is consistent with the existence of the scaled
quantities
\begin{equation}\label{sle}
a(s,y) := \lim_{t \to \infty} P_{[2t + t^{1/3} s]}(1 - y/t^{1/3}), \qquad
b(s,y) :=  \lim_{t \to \infty} Q_{[2t + t^{1/3} s]}(1 - y/t^{1/3}),
\end{equation}
and that furthermore (\ref{13.0}) reduces to the partial differential equations
\begin{equation}\label{3.36}
{\partial a \over \partial s} = q b, \qquad
{\partial b \over \partial s} = q a - yb,
\end{equation}
where use has also been made of (\ref{j.3}) and (\ref{13.3}).
We note too that existence of the limits (\ref{sle}) together with the
formula
$$
\tilde{D}_l \Big |_{\alpha_+=1/\alpha_- = \alpha}/D_l =
{1 \over \kappa_l^2} \sum_{k=0}^{l} \kappa_k^2 \pi_k(\alpha_+)
\pi_k(1/\alpha_-),
$$
which follows from the final equality in (\ref{6.41}), allows for the 
formal derivation of the limit
\begin{eqnarray}\label{6.62a}
&&g(s,y):= \lim_{t \to \infty} e^{-(\alpha_+ + \alpha_-)t}
\tilde{D}_{[2t + t^{1/3} s]}(t) \Big |_{\alpha_+ = 1/\alpha_- = 1 -
y/t^{1/3}}/D_{[2t + t^{1/3} s]} \nonumber \\
&& \qquad = \int_{-\infty}^s a(s',y) a(s',-y) \, ds' =
 \int_{-\infty}^s b(s',y) b(s',-y) \, ds'
\end{eqnarray}
Similarly, at the same formal level, we see from (\ref{4.41}), the definition
of $g(s,y)$, and (\ref{4.47}) that
\begin{eqnarray}\label{4.67a}
\tilde{F}_y(s) & := & \lim_{t \to \infty}
{\rm Pr} \Big ( {L(t,1 - y/t^{1/3}, 1+ y/t^{1/3}) - 2 \sqrt{t} \over
t^{1/6} } \le s \Big ) \nonumber \\
& = & {\partial \over \partial s} \Big ( g(s,y) F_{\rm GUE}(s) \Big ).
\end{eqnarray} 

To fully determine $a$ and $b$, and thus $g(s,y)$ and the scaled distribution
$\tilde{F}_y(s)$,
it remains to specify equations for their dependence on the
scaled variable $y$ as  introduced in (\ref{ay}). Such equations follow from
differential equations in $z$ for $\pi_n$ and $\pi^*_n$ \cite{IW01,Ba01,PS02}.

\begin{prop}
We have
\begin{eqnarray}
\pi_n'(z) & = & \Big ( {n \over z} + {t \over z^2} - {r_{n+1} r_n t \over
z } \Big ) \pi_n(z) + \Big ( {r_{n+1} t \over z} - {r_n t \over z^2}
\Big ) \pi_n^*(z) \label{iw.1} \\
{\pi_n^{*}}'(z) & = &  \Big ( - {r_{n+1} t \over z} + r_n t \Big )
\pi_n(z) + \Big ( -t + {r_{n+1} r_n t \over z} \Big ) \pi_n^*(z).
\label{iw.2}
\end{eqnarray}
\end{prop}

\noindent
Proof. \quad Let us show how (\ref{iw.1}) can be derived using results from
\cite{IW01}. In terms of the polynomials (\ref{11.d4}) for the weight
(\ref{rt.1}), it is shown in \cite[eq.~(2.79) with $t \to 2t$]{IW01}
\begin{eqnarray}\label{ter}
\phi_n'(z) & = & {\kappa_{n-1} \over \kappa_n} \Big (
n + {t \over z} + t {\kappa_{n-1} \phi_{n-1}(0) \over \kappa_n \phi_n(0)} -
t {\phi_{n+1}(0)\phi_n(0) \over \kappa_{n+1} \kappa_n} \Big ) \phi_{n-1}(z)
\nonumber \\
 &  & - {t \over z} {\kappa_{n-1} \over \kappa_n} {\phi_{n-1}(0) \over
\phi_n(0)} \phi_n(z).
\end{eqnarray}
But for a general weight \cite{Sz75}
$$
\kappa_{n-1} z \phi_{n-1}(z) = \kappa_n \phi_n(z) - \phi_n(0) \phi_n^*(z),
\qquad 1 - r_n^2 = \Big ( {\kappa_{n-1} \over \kappa_n} \Big )^2 , 
$$ 
which together with (\ref{rt.2}) show (\ref{ter}) reduces to (\ref{iw.1}).
\hfill $\square$

The differential equations (\ref{iw.1}), (\ref{iw.2})
are consistent with the existence of the limits
(\ref{sle}), and furthermore assume the scaled form
\begin{equation}\label{3.37}
{\partial \over \partial y} a = q^2 a - (q' + y q)b, \qquad
{\partial \over \partial y} b = (q' - yq)a + (y^2 - s - q^2)b.
\end{equation}
Together (\ref{3.36}) and (\ref{3.37}) determine the scaled quantities
(\ref{sle}) once appropriate initial conditions are specified.

According to (\ref{ay}), when $y=0$, $\alpha=1$, so according to (\ref{sle})
we require the behaviour of $\pi_n(-1)$, $\pi_n^*(-1)$. First we note from
(\ref{6.40'})  that $\pi_n^*(-1) = (-1)^n \pi_n(-1)$ and thus
(\ref{sle}) gives $a(s,0) = - b(s,0)$. It then follows from (\ref{3.36}) that
\begin{equation}
a(s,0) = A e^{-U(s)}, \qquad U(s) = - \int_s^\infty q(t) \, dt.
\end{equation}
To determine $A$, we note that formulas in \cite{BR01a} imply
$$
\lim_{n \to \infty} (-1)^n \pi_n(-1) = e^t
$$
so we see from (\ref{sle1}) and (\ref{sle}) that
$$
a(s,0) \to 1 \qquad {\rm as} \qquad s \to \infty
$$
and so
\begin{equation}\label{ic}
a(s,0) = - b(s,0) = e^{-U(s)}.
\end{equation}
The quantities $a(s,y)$, $b(s,y)$ are now fully determined, and thus so to
is $g(s,y)$ as specified by (\ref{6.62a}) and the scaled cumulative 
distribution $\tilde{F}_y(s)$ as specified by (\ref{4.67a}).

The scaled quantities satisfy a number of further properties of interest.
First,
with the initial condition (\ref{ic}) it is easy to see from (\ref{3.36})
and (\ref{3.37}) that $a(s,y)$ and $b(s,y)$ are related by
\begin{equation}\label{4.74}
a(s,y) = - b(s,-y) e^{{1 \over 3} y^3 - sy}.
\end{equation}
From this it is simple to verify that (\ref{6.62a}) can alternatively be
written \cite{BR01a}
\begin{equation}\label{6.62b}
g(s,y) = a(s,-y) {\partial \over \partial y} a(s,y) - 
b(s,-y) {\partial \over \partial y} b(s,y),
\end{equation}
which relates to (\ref{D.D}). 
Second, it follows from (\ref{11.7}), (\ref{3.36}) and (\ref{ic}) that
$$
b(s,y) \mathop{\sim}\limits_{s \to \infty} - e^{-ys}.
$$
One then sees from (\ref{4.74}), (\ref{3.37}) and (\ref{6.62b}) that
\begin{equation}\label{4.75a}
g(s,y)  \mathop{\sim}\limits_{s \to \infty} s - y^2.
\end{equation}
Using this, the mean of the distribution $d \tilde{F}_y/ ds$ can be
computed as
$$
\int_{-\infty}^\infty s \, d \tilde{F}_y =
\lim_{s \to \infty} \Big ( s \tilde{F}_y(s) - g(s,y) F_{\rm GUE}(s)
\Big ) = y^2
$$
where the first equality follows from integration by parts and (\ref{4.67a}),
while the second equality follows from the facts that
$\tilde{F}_y(s)$ and $F_{\rm GUE}(s)$ approach 1 exponentially fast,
together with (\ref{4.75a}). This then motivates defining the
shifted cumulative distibution \cite{PS02}
$$
F_y(s) :=  \tilde{F}_y(s+y^2)
$$
for which the corresponding distribution function $d F_y/ ds$ has mean zero.
From the discussion in \cite{PS02}, ones sees that
of immediate interest to KPZ growth is the second moment
$$
\int_{-\infty}^\infty s^2 \, dF_y,
$$
which is a function of the scaled parameter $y$. Pr\"ahofer and Spohn 
\cite{PS02} have
used the results revised above, evaluated using high precision computing,
to accurately tabulate this quantity and discuss its properties.

\section{Scaled limits for the symmetrized models}
In a remarkable analysis Baik and Rains \cite{BR01b} have provided similar
evaluations to the one detailed in Section 4 of the scaled limits
of the cumulative distributions in Section 3. Here we will be content
with drawing attention to a subcase of two of these:
${\rm Pr}(l^\symmO \le l )$ in the case $\alpha = 0$ and
${\rm Pr}(l^\symmS \le 2l )$ in the case $\beta = 0$. Thus
it was first proved
in  \cite{BR01b} (see \cite{BF03} for a subsequent simplified derivation based
on the identities (\ref{j.O}) and (\ref{j.Oa})) that
\begin{eqnarray*}
\lim_{z \to \infty} {\rm Pr} \Big ( {l^\symmO(z) - 2 z \over z^{1/3} }
\le s \Big ) & = & F_{\rm GSE}(s) \\
\lim_{z \to \infty} {\rm Pr} \Big ( {l^\symmS(z) - 2 z \over z^{1/3} }
\le s \Big ) & = & F_{\rm GOE}(s).
\end{eqnarray*}
Here $F_{\rm GSE}(s)$ denotes the scaled cumulative distribution of the
largest eigenvalue for large Hermitian random matrices with real
quaternion elements, while $F_{\rm GOE}(s)$ denotes the same for large
real symmetric matrices.

Both $F_{\rm GSE}(s)$ and $F_{\rm GOE}(s)$ can be expressed in terms of
the same Painlev\'e II transcendent as in the evaluation (\ref{4.47a}) of
$F_{\rm GUE}(s)$. Thus one has \cite{TW96}
\begin{eqnarray*}
F_{\rm GSE}(s) & = & {1 \over 2} \Big ( F_{\rm GUE}(s) \Big )^{1/2}
\Big ( e^{{1 \over 2} \int_s^\infty q(t) \, dt} +
 e^{-{1 \over 2} \int_s^\infty q(t) \, dt} \Big ) \\
F_{\rm GOE}(s) & = &  \Big ( F_{\rm GUE}(s) \Big )^{1/2}
e^{{1 \over 2} \int_s^\infty q(t) \, dt}.
\end{eqnarray*}
Note the obvious inter-relationship
$$
F_{\rm GSE}(s) = {1 \over 2} \Big ( F_{\rm GOE}(s) + {F_{\rm GUE}(s) \over
F_{\rm GOE}(s)} \Big ). 
$$
The origin of such an identity can be traced back to a special property
of a particular marginal distribution of the joint probability for the
row lengths of semi-standard tableaux relating to the symmetry
$\symmS$ \cite{BR01a,FR01,FR02b}. 
The marginal distribution is defined by summing
over every second row of the semi-standard tableaux. We also draw
attention to the fact that $F_{\rm GUE}(s)$ and $F_{\rm GOE}(s)$ are
$\tau$-functions for certain Painlev\'e II systems \cite{TW94a,FW01}.
Similarly, $F_{\rm GSE}(s)$ is the arithmetic mean of two
$\tau$-functions, both of which correspond to Hamiltonians satisfying
the same differential equation, differing only in the boundary
condition \cite{FW01}.


\end{document}